\documentclass[sigconf]{acmart}
\usepackage{siunitx}
\usepackage{bm} 
\AtBeginDocument{%
  }

\copyrightyear{2026}
\acmYear{2026}
\setcopyright{cc}
\setcctype{by}
\acmConference[ICMR '26]{International Conference on Multimedia Retrieval}{June 16--19, 2026}{Amsterdam, Netherlands}
\acmBooktitle{International Conference on Multimedia Retrieval (ICMR '26), June 16--19, 2026, Amsterdam, Netherlands}
\acmDOI{10.1145/3805622.3810434}
\acmISBN{979-8-4007-2617-0/2026/06}

\begin{document}

\title{Realistic Virtual Flood Experience System Using 360° Videos and \\3D City Models Constructed from Building Footprints}

\author{Tatsuro Banno}
\affiliation{%
  \institution{The University of Tokyo}
  \city{Tokyo}
  \country{Japan}
}
\email{banno@hal.t.u-tokyo.ac.jp}

\author{Koki Kawada}
\affiliation{%
  \institution{The University of Tokyo}
  \city{Tokyo}
  \country{Japan}
}
\email{kawada@hal.t.u-tokyo.ac.jp}

\author{Mizuki Takenawa}
\affiliation{%
  \institution{The University of Tokyo}
  \city{Tokyo}
  \country{Japan}
}
\email{takenawa@hal.t.u-tokyo.ac.jp}

\author{Masatoshi Denda}
\affiliation{%
  \institution{Public Works Research Institute}
  \city{Tsukuba}
  \country{Japan}
}
\email{denda-m933jp@pwri.go.jp}

\author{Kiyoharu Aizawa}
\affiliation{%
  \institution{The University of Tokyo}
  \city{Tokyo}
  \country{Japan}
}
\email{aizawa@hal.t.u-tokyo.ac.jp}

\renewcommand{\shortauthors}{Banno et al.}

\newcommand{\ie}{{\it i.e.}}
\newcommand{\eg}{{\it e.g.}}
\newcommand{\etal}{{\it et al. }}

\begin{abstract}
Virtual flood experience systems, which enable users to vividly experience flooding, are attracting increasing attention as effective tools for communicating flood risks. However, existing systems typically rely on virtual cities that do not correspond to real locations and often lack sufficient photorealism, limiting users’ ability to relate scenarios to their own surroundings. Although 360° video-based virtual environments offer a simple and scalable way to visually replicate real-world scenes, effective 3D flood visualization in these environments typically requires 3D building geometry of the target area, which is not readily available in many regions. To address this limitation, we propose a new virtual flood experience framework that integrates 360° videos with 3D models automatically constructed from widely available 2D building footprints. By extruding footprints to plausible heights and spatially aligning the constructed models with 360° videos, our framework enables 3D flood visualization in photorealistic environments without relying on pre-existing city models such as CityGML. We demonstrate the framework in Memuro, Hokkaido, Japan, an area vulnerable to river flooding. A user study with local residents showed that the proposed system enhances users’ ability to envision location-specific flood evacuation situations, demonstrating its potential as an effective tool for disaster risk communication and education.
\end{abstract}

\ccsdesc[500]{Information systems~Multimedia content creation}
\ccsdesc[500]{Information systems~Geographic information systems}

\keywords{disaster awareness, 360° video, urban visualization}
\begin{teaserfigure}
\centering
  \includegraphics[width=0.97\textwidth]{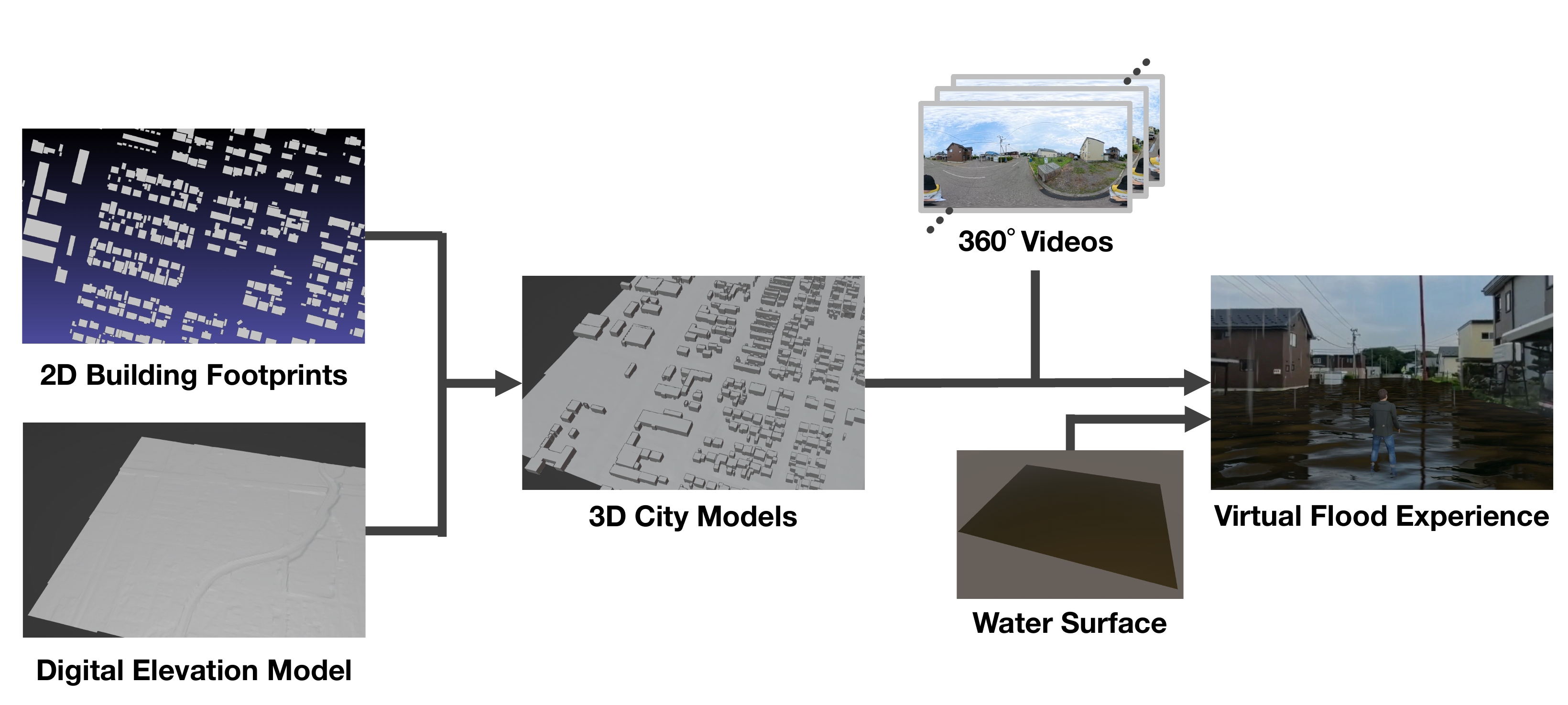}
  \caption{An overview of our proposed virtual flood experience system that integrates 360° videos with 3D city models automatically generated from 2D building footprints, enabling 3D flood visualizations in photorelistic environment without relying on pre-existing city models such as CityGML.}
  \label{fig:teaser}
\end{teaserfigure}

\maketitle

\section{Introduction}

Floods are devastating worldwide disasters; however, many local residents underestimate their impacts, leading to insufficient preparedness~\cite{burningham2008ll}. To address this awareness gap, virtual flood experience systems, which enable people to vividly experience flooding in virtual environments, are attracting increasing attention as risk communication tools. Recently, numerous studies have developed and evaluated virtual flood experience systems, implemented using head-mounted displays~\cite{sermet2019flood,fujimi2020testing} or desktop-based platforms~\cite{d2023non, denda2024development}.

Existing virtual flood experience systems have limitations in the realism of their virtual environments. In many cases, such systems are implemented in purely virtual cities that do not correspond to actual locations~\cite{fujimi2020testing, d2023non}, making it difficult for residents to relate the scenarios to their own environment. Some approaches generate virtual environments of real cities using procedural modeling based on real-world data~\cite{sermet2019flood}, but they often lack sufficient photorealism and fail to capture true urban appearances. More recently, Denda \etal explored highly detailed 3D environments of real cities using CAD models and photogrammetry~\cite{denda2024development}. While these methods improve realism, they require extensive data collection and labor-intensive processing, making it difficult to scale them to larger areas or multiple cities.

In contrast, recent studies have shown that 360° walkthrough videos recorded along streets provide a simple and scalable way to generate photorealistic virtual environments of a target area~\cite{sugimoto2020urban, sugimoto2020building, takenawa2023360rvw, takenawa2026building, banno2026360citygml}. By capturing the entire surroundings simutaneously, 360° videos allow users to freely change 
their viewing direction, producing realistic and immersive representations of the environment without labor-intensive 3D modeling. Building on this concept, Movie Map~\cite{sugimoto2020urban, sugimoto2020building} enables continuous exploration of the target area by automatically detecting intersections along routes using Visual SLAM~\cite{sumikura2019openvslam} and seamlessly transitioning between video segments. These systems have also been extended to incorporate user-controlled avatars~\cite{takenawa2023360rvw, takenawa2026building}, supporting highly interactive and immersive navigation of the environments.

More recently, Banno \etal~\cite{banno2026360citygml} proposed a novel urban visualization system that integrates 360° video-based virtual environments with 3D urban models (\ie, CityGML~\cite{groger2012citygml}). By aligning video frames with the 3D models and dynamically projecting them onto the geometries, this approach enables realistic visualization of 3D geospatial data (\eg, floodwater levels or daylight shadows) within photorealistic representations of real-world environments. While applying this method to virtual flood experience systems represents a promising direction, it depends on 3D building geometry that is often unavailable in many regions, limiting its applicability.

To address this limitation, we propose a new virtual flood experience framework that integrates 360° videos with 3D models automatically generated from widely available 2D building footprints. Our key insight is that for supporting virtual 3D flood visualization, precise building heights are not required; it is sufficient to model the lower portions of buildings that interact with floodwaters. Based on this insight, we first generate 3D building models by extruding footprints with constant heights (\eg, 20\,\si{\meter}), producing simplified but flood-relevant geometries. Next, we spatially align the generated models with 360° videos. Unlike existing building-region-based alignment methods~\cite{banno2026360citygml} which assume accurate building heights, our approach relies on ground regions and 3D points obtained via Visual SLAM, eliminating the need for precise pre-existing 3D city models. Finally, we use these aligned geometries and videos to dynamically project relevant video frames onto the models~\cite{banno2026360citygml}, enabling realistic 3D flood visualization in real-world environments without relying on detailed pre-existing 3D city models.

Using the proposed framework, we demonstrate a virtual flood experience system in Memuro, Hokkaido, Japan, an area vulnerable to river flooding. To evaluate the effectiveness of our system, we organized a workshop with local residents and conducted a user study on their experience. The results show that it enhances users’ ability to envision location-specific flood evacuation scenarios, highlighting its potential as an effective tool for disaster risk communication and education.

\section{System implementation}

An overview of the system is shown in Fig.~\ref{fig:teaser}. Our system uses two main inputs: 1) 2D Building footprints and Digital Elevation Model (DEM) of the target area. These datasets provide the geometric foundation for constructing the 3D virtual environment, as detailed in Sec.~\ref{sec:construction}. Unlike detailed urban 3D models such as CityGML~\cite{groger2012citygml}, 2D building footprints and DEM data are widely available across large regions. For example, in Japan, datasets provided by the Geospatial Information Authority of Japan cover approximately $98\,\%$ of the country~\cite{mlit2026}, enabling nationwide deployment of our approach. 2) 360° walkthrough
videos of the target area. Following previous virtual systems~\cite{sugimoto2020urban, sugimoto2020building, takenawa2023360rvw, takenawa2026building, banno2026360citygml}, we use multiple 360° walkthrough videos to represent the target area, with each captured along a single street segment in a street-by-street manner. The start and end points of each video are manually annotated on a map during data collection. 

From these inputs, we first generate 3D building models by extruding the footprints (Sec.~\ref{sec:construction}). We then align the models with 360° videos (Sec.~\ref{sec:alignment}) and finally project videos onto the models~\cite{banno2026360citygml} to create a realistic and interactive urban flood experience system in which users navigate the environment using avatars (Sec.~\ref{sec:integration}). The following sections describe the technical details of each step.

\subsection{Construction of 3D city models}
\label{sec:construction}

To provide a geometric foundation for 3D flood visualization in our system, we first construct 3D city models of the target area from the building footprints and DEM. We use the DEM directly as the terrain surface by treating its gridded elevation values (\eg, 5\,\si{\meter} spacing) as ground heights. Starting from this terrain surface, we vertically extrude the corresponding 2D building footprints to a constant height of 20\,\si{\meter}, generating simplified 3D building models for flood visualization. Although these models do not represent the actual building heights, they accurately capture the lower portions of the buildings that interact with floodwater, providing a sufficient foundation for 3D flood visualization.

\subsection{Alignment of 3D models and 360° videos}
\label{sec:alignment}

\begin{figure}[t]
    \centering
    \includegraphics[width=\linewidth]{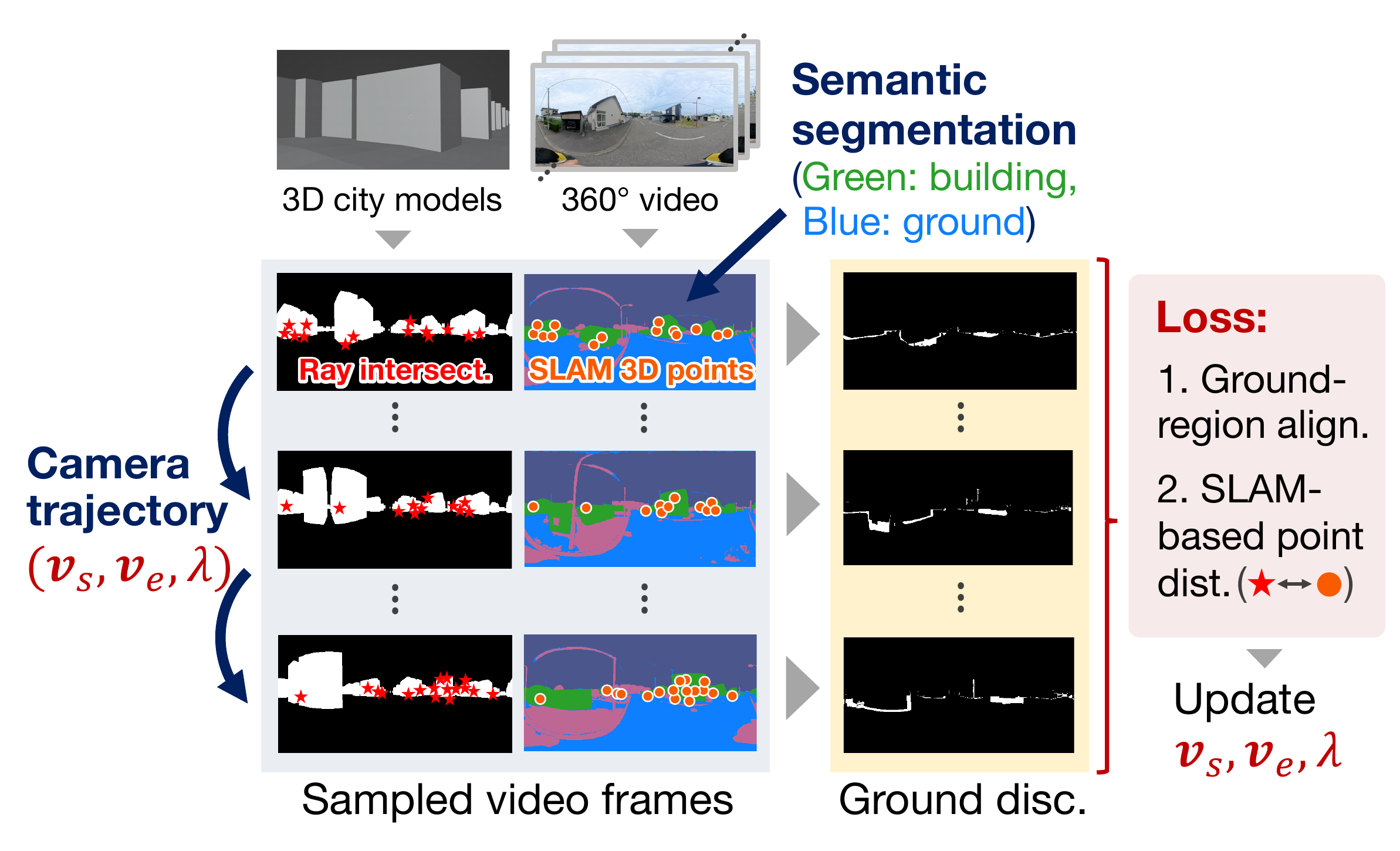}
    \caption{An overview of our optimization process for aligning 3D city models with 360° videos, using ground-region alignment loss and SLAM-based point distance loss to refine parameters.}
    \label{fig:opt_overview}
\end{figure}
Next, to accurately integrate the 360° videos with the constructed 3D city models, we align the models with the video frames. Following~\cite{banno2026360citygml}, we formulate this task as a street-level alignment problem. Specifically, we first run Visual SLAM~\cite{sumikura2019openvslam} on each 360° walkthrough video to estimate camera trajectories in a local coordinate system. We then register the local trajectories to the world coordinate system of the 3D models by optimizing a transformation between the local camera coordinate system and the world coordinate system. In practice, we optimize the 3D positions of the trajectory’s start and end points in the world coordinate system, $\bm{v}_s=(x_s, y_s, z_s)$ and $\bm{v}_e=(x_e, y_e, z_e)$, as well as the rotation around the gravity axis, $\lambda$, which together fully determine the transformation matrix.

In previous work~\cite{banno2026360citygml}, these parameters were optimized by minimizing the discrepancy between building regions in the 3D models and 360° video frames. In our setting, however, building-region-based alignment is unreliable because the heights of 3D models are uniform and do not reflect actual building geometries. We therefore propose a new alignment method that does not rely on building heights, instead leveraging discrepancies in ground regions and 3D points obtained from Visual SLAM.

Fig.~\ref{fig:opt_overview} illustrates an overview of our proposed optimization process. To reduce computational cost, we sample a small number of frames from the 360° video and use them for optimization. For each selected frame, we first use a semantic segmentation model~\cite{cheng2022masked} trained on a city landscape dataset~\cite{neuhold2017mapillary} to extract the building and ground regions. We then compute the following two losses to guide the optimization, aiming to minimize the misalignment between the 3D models and the 360° video frames.
\\

\noindent\textbf{1. Ground-region alignment loss:}  
To overcome the limitations of building-region-based alignment~\cite{banno2026360citygml} which depends on accurate building heights, we leverage the ground region in 360° video frames as a stable reference for alignment. Specifically, for each sampled frame, we first project the constructed terrain and building models onto the frame using the estimated camera pose. We then compute the pixel-wise discrepancy between the ground regions in the video and the projection, and use the sum of these discrepancies as the ground-region loss for optimization.
\\

\noindent\textbf{2. SLAM-based point distance loss:}
After the SLAM process introduced earlier in the section, we obtain 3D coordinates of feature points, each linked to the keyframe that produced it. To provide additional supervision for alignment, we incorporate these feature points into the optimization process. First, to focus on building points, we reproject each 3D point onto its corresponding camera pose in the local SLAM coordinate system and retain only those within the building regions of the corresponding video frame. Next, for each retained point, we cast a ray from the camera center toward the point and compute the distance at which it intersects the 3D model. Finally, we calculate the differences between these ray intersection distances and the camera-to-point distances obtained from SLAM, and define their sum as the SLAM-based point distance loss. This loss captures misalignments between the reconstructed model and the original SLAM observations, providing dense supervision for alignment with the observed scene.
\\

\begin{figure}[t]
    \centering
  \includegraphics[width=\linewidth]{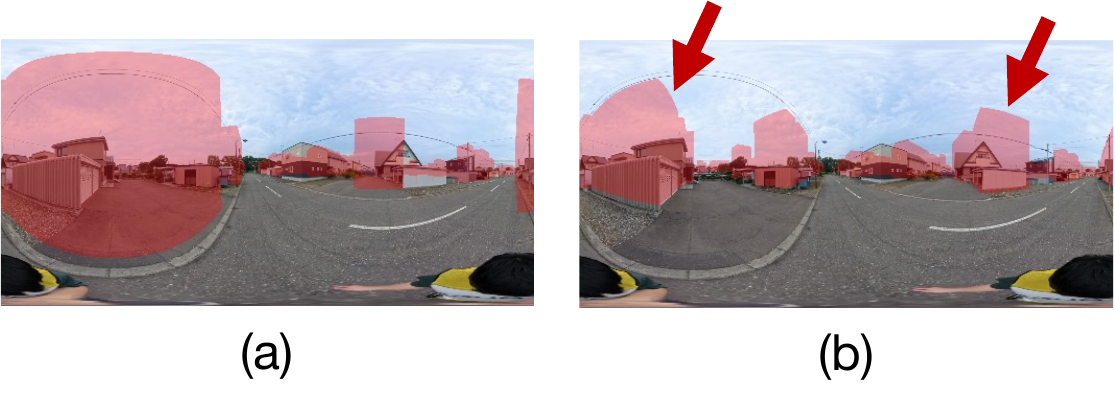}
    \caption{Comparison of discrepancies between 360° video frames and the 3D city model. (a) before optimization and (b) after optimization. Red-highlighted areas indicate the projections of the 3D building models onto the 360° camera view, based on the estimated camera poses. Note that the building heights are set uniform and inaccurate; pay attention to the boundaries between the buildings and the ground.}
    \label{fig:opt_result}
\end{figure}

Using these losses, we iteratively update the optimization parameters $\bm{v}_s$, $\bm{v}_e$, and $\lambda$ via Covariance Matrix Adaptation Evolution Strategy (CMA-ES)~\cite{hansen2001completely}. Fig.~\ref{fig:opt_result} shows qualitative results before and after the optimization process. Compared with the initial alignment, our method successfully aligns each 360° frame with the 3D models, even though building heights are inaccurate. This demonstrates the effectiveness of our alignment approach.

\subsection{Integration and flood visualization}
\label{sec:integration}

Finally, given aligned 3D city models and 360° videos, we create a photorealistic virtual environment in which users navigate using avatars. Following~\cite{banno2026360citygml}, we apply texture projection of 360° video frames onto the 3D urban model geometries. As users move their avatars, the system selects the optimal camera position along the camera trajectory and projects the corresponding 360° video frame onto the 3D geometry. By continuously updating the camera and projected frames based on the avatar’s movement, the system enables free navigation within a photorealistic, video-based environment.

\begin{figure}[t]
    \centering
    \includegraphics[width=\linewidth]{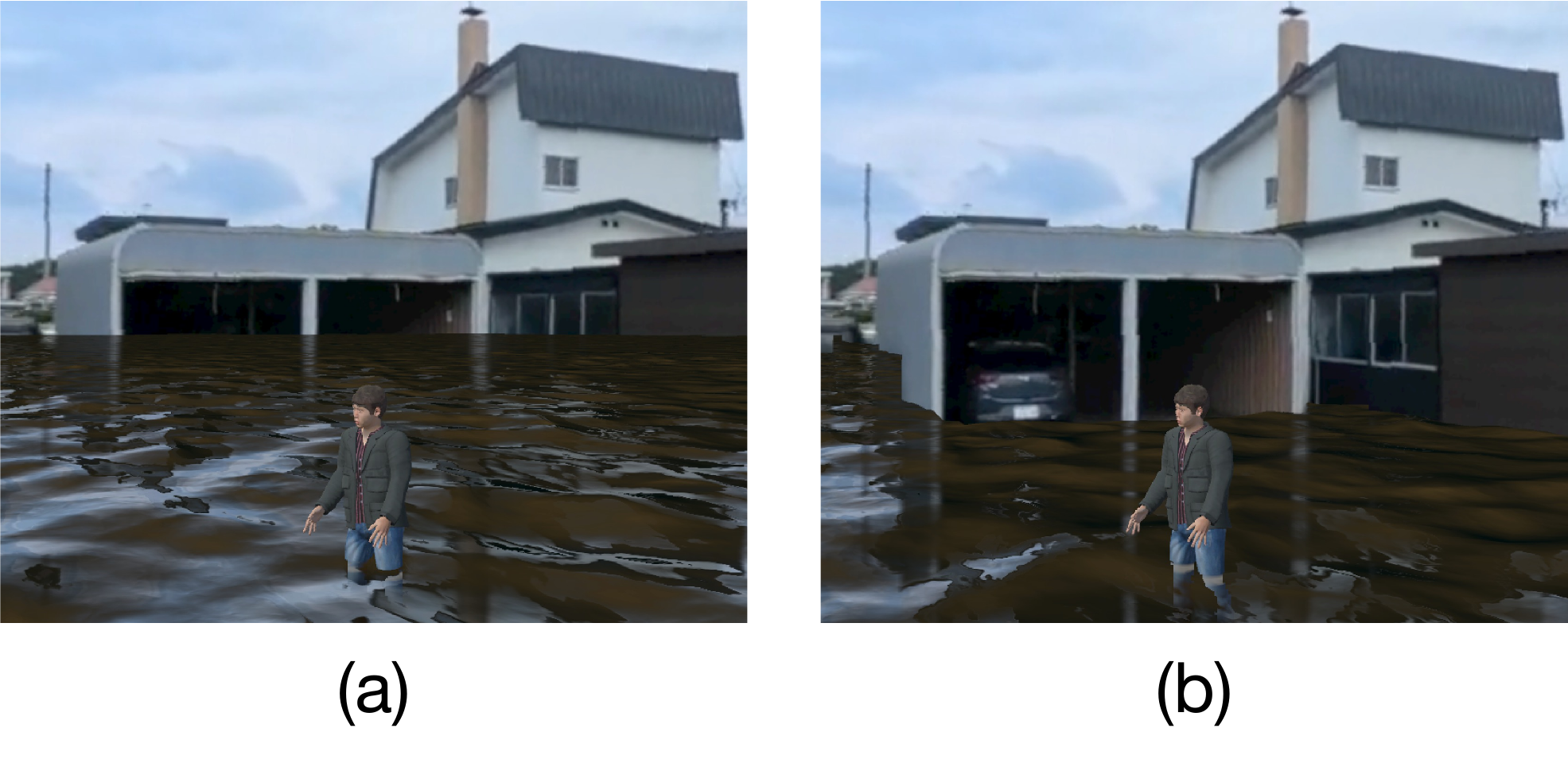}
    \caption{Comparison between a flood visualization in (a) a 360° video-based environment without 3D geometry and (b) our virtual environment, which incorporates 3D geometry reconstructed from 2D building footprints.}
    \label{fig:comparison}
\end{figure}

Moreover, by aligning 3D urban models with 360° videos, our virtual environment enables realistic 3D flood visualizations. Fig.~\ref{fig:comparison} compares a flood visualization in a 360° video-based environment without 3D geometry~\cite{takenawa2023360rvw, takenawa2026building} to our system, which incorporates 3D geometry reconstructed from 2D building footprints. While discrepancies between building façades and floodwater are evident in the video-only environment, incorporating 3D geometry allows flood visualizations to more accurately represent the urban structure. This highlights the importance of integrating 3D geometry in 360° video-based virtual flood experience systems.

\section{Demonstration}

Using the proposed framework, we developed a virtual flood experience system in Memuro, Hokkaido, Japan, using Unity~\cite{unity}\footnote{Demo video is available at \url{https://youtu.be/9IvnP2ndzL0}}
. The area suffered significant flood damage in 2016~\cite{memuro2017} and is considered at high risk for future flood disasters. In this area, we captured 360° video and utilized 2D building footprints and DEM data provided by the Geospatial Information Authority of Japan~\cite{kiban}.

Focusing on Memuro, we implemented a virtual flood experience that simulates actual flood evacuation scenarios. Fig.~\ref{fig:demo} shows example views of the system. In the system, users are instructed to move to predefined evacuation points (\eg, Memuro Elementary School) by navigating an avatar through the virtual environment (Fig.~\ref{fig:demo} (a)). If a user fails to reach an evacuation point in time, the flood surface appears and the avatar is overtaken by flood water (Fig.~\ref{fig:demo} (b)), emphasizing the importance of timely evacuation.

\section{User study}

To evaluate our system, we organized a workshop in Memuro, Hokkaido, Japan, inviting local residents to participate in a user study. A total of 29 residents attended, and due to time constraints, they were divided into 12 groups of 2–5 people each. During the workshop, each group watched an instructor operate the demonstration, and afterward, each group completed a single questionnaire to provide feedback on their experience.

We asked participants two questions: Q1: whether they clearly understood their location in the town during the demonstration, and Q2: whether the system helped them better imagine actual flood evacuation scenarios. Participants responded using a 5-point Likert scale (1 = lowest, 5 = highest). For Q1, the average score was 4.92 (SD = 0.29), indicating that users could easily recognize their real-world location within the virtual environment, which is an advantage over systems that use purely virtual cities with no correspondence to actual locations~\cite{fujimi2020testing, d2023non}. For Q2, the average score was 4.75 (SD = 0.62), suggesting that the proposed system also enhances users’ ability to envision location-specific flood evacuation scenarios. Overall, these results highlight the system’s potential as an effective tool for disaster risk communication and education.

\begin{figure}[t]
    \centering
    \includegraphics[width=\linewidth]{fig/demo.pdf}
    \caption{Example views of our virtual flood experience system in Memuro, Hokkaido, Japan. (a) Users navigate an avatar through the virtual environment to reach predefined evacuation points. (b) If a user fails to reach an evacuation point in time, the flood surface rises and overtakes the avatar.}
    \label{fig:demo}
\end{figure}

\section{Conclusion}

We present a novel virtual flood experience framework that integrates 360° videos with 3D models automatically generated from widely available 2D building footprints. By extruding footprints to plausible heights and aligning the resulting models with 360° videos, our framework enables photorealistic 3D flood visualizations without relying on pre-existing 3D city models. We demonstrate the framework in Memuro, Hokkaido, Japan, and evaluate it through a user study with local residents. The study suggests that the system can support participants in better understanding and envisioning location-specific flood evacuation scenarios. We hope that our framework may serve as a practical tool for flood risk communication and education, supporting improved flood preparedness among local residents.

\begin{acks}
This research was partly supported by JSPS KAKENHI 25H01164 and CSTI SIP Resilient Smart Network System against Natural Disasters.
\end{acks}

\bibliographystyle{ACM-Reference-Format}
\bibliography{main-base}

\end{document}